\documentclass[twocolumn,showpacs,preprintnumbers,prl,amsmath,amssymb]{revtex4}

\usepackage{graphicx}
\usepackage{dcolumn}
\usepackage{bm}

\begin{document}


\title{Thermally excited capillary waves at vapor/liquid interfaces of water-alcohol mixtures}
\author{David Vaknin and Wei Bu}
\affiliation{Ames Laboratory and Department of Physics and Astronomy, Iowa State University, Ames, Iowa 50011, U.S.A.}
\author{Jaeho Sung, Yoonnam Jeon, and Doseok Kim}
\affiliation{Department of Physics and Interdisciplinary Program of Integrated Biotechnology, Sogang University, Seoul 121-742, Korea}
\date{\today}

\begin{abstract}
The density profiles of liquid/vapor interfaces of water-alcohol (methanol, ethanol and propanol)
mixtures were studied by surface sensitive synchrotron X-ray scattering techniques.  X-ray reflectivity and diffuse scattering measurements, from the pure and mixed liquids, were analyzed in the framework of capillary-wave theory to address the characteristics length-scales of the intrinsic roughness and the shortest capillary-wavelength (alternatively, the upper wave-vector cutoff in capillary wave theory).  Our results establish that the intrinsic roughness is dominated by average interatomic distances.  The extracted effective upper wave-vector cutoff indicates capillary wave theory breaks-down at distances on order of bulk correlation lengths.
\end{abstract}

\maketitle
\section{Introduction}
It is by now common theoretically
\cite{Cahn1958,Fisk1969,Buff1965,Weeks1977,Rowlinson1982,Beysens1987,Meunier1987,Sengers1989,Gelfand1990}
and experimentally
\cite{Huang1969,Wu1973,Braslau1988,Daillant1989,Schwartz1990,Sanyal1991,Ocko1994,Fukuto1998,Pershan2000,Shpyrko2004,Fukuto2006}
that the density profile of simple liquid/vapor interfaces is
dominated by thermally excited capillary waves.  Theoretically,
the profile is derived by statistical mechanics tools assuming
thermal excitations in a two-dimensional membrane under surface
tension and under the influence of gravity on mass displacements.
Initially, the density profiles were measured in the vicinity of
the critical point of liquids, where the interfacial profile is sufficiently wide
(i.e., very low surface tension) to be adequately resolved by
light scattering techniques \cite{Huang1969,Wu1973}.
However, with the advent in X-ray scattering technique for liquid
surfaces \cite{Als-Nielsen1983}, the profiles of common simple
liquids, such as water and alcohols, away from the critical point
have been determined to be a few Angstrom thick
\cite{Braslau1988}.

\begin{figure}[thl]
\includegraphics[width=3.0 in, angle=0]{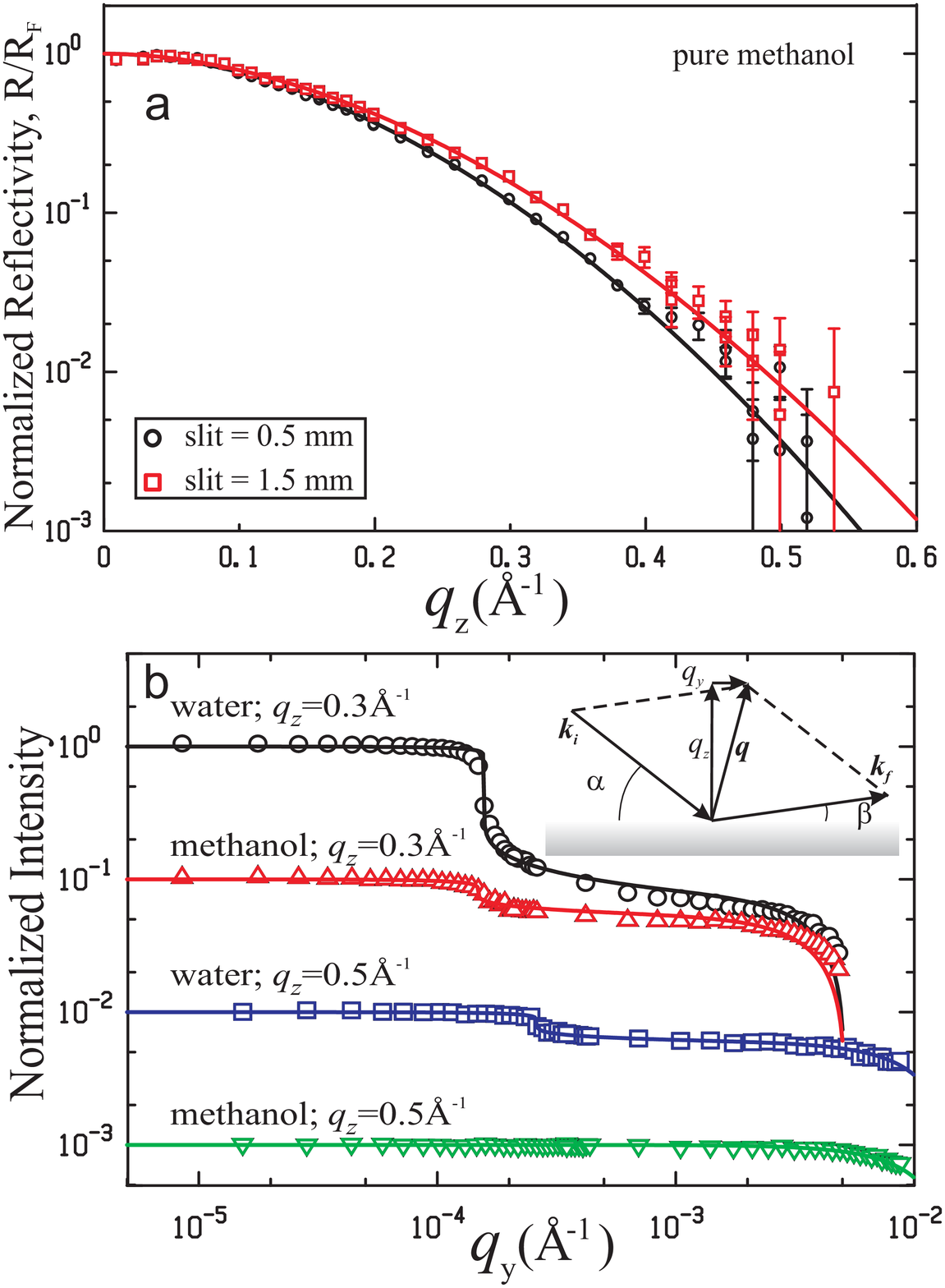}
\caption{\label{Ref0} (a) (color online) Normalized X-ray reflectivities of pure
methanol {\it versus} $q_z$ demonstrating the effect of the
resolution on the measurement (detector slit sizes indicated).
The solid lines were obtained by simultaneous fitting using Eq.\
(\ref{For-RRF}). (b) The XDS from pure water and pure methanol at
constant $q_z$ values as indicated (shifted by a decade each for clarity). The solid lines are fits to
the data using Eq.\ (\ref{For-XDS}). The inset shows the
experimental beam geometry.}
\end{figure}

The X-ray reflectivity from a liquid surface $R(q_z)$ ($q_z =
2k_0\sin\alpha$; $\alpha$ incident beam angle,
$k_0=2\pi/\lambda$; $\lambda$ X-ray wavelength; see inset in fig.\
\ref{Ref0})  is given by
\cite{Braslau1988,Sinha1988,Sanyal1991,Fukuto2006,BuTobepublished}
\begin{equation}
R(q_z)=R_F(q_z)R(0,q_z){\rm{e}}^{-\sigma_0^2q_z^2}
\label{For-Rqz}
\end{equation}
where $R_F(q_z)$ is the Fresnel reflectivity from an ideally flat
surface, $\sigma_0$ is the intrinsic roughness, and for a
rectangular-shaped resolution function (used in the present
experiment) \cite{Sinha1988}
\begin{equation}
R(0,q_{z}) =\frac{2^{1-2\eta}\Gamma(1/2-\eta/2)}{\sqrt{\pi}\eta\Gamma(\eta/2)}\left(\frac{\Delta q_{y}}{q_{\max}}\right)^{\eta}
\label{For-R0qz}
\end{equation}
where $\eta =\frac{k_BT}{2\pi\gamma}q_z^2$ and $\Delta
q_{y}=q_{z}\Delta\beta/2$. Here, $\Delta\beta$ is the detector
acceptance angle, $\gamma$ is a surface tension, and $T$ is the
sample temperature. In the $q_z$-range of a typical
reflectivity measurement, $\eta$ values are small and we can use
the following expression,
\begin{equation}
R(q_z)/R_F\approx{\rm{e}}^{(-\sigma_{cw}^{2}-\sigma_0^2)q_z^2} =
{\rm{e}}^{-\sigma_{\rm{eff}}^2 q_z^2}, \label{For-RRF}
\end{equation}
where the effective roughness can be written as follows
\cite{Braslau1988,Schwartz1990,Ocko1994,Pershan2000,Fukuto2006}.
\begin{equation}
\sigma_{\rm{eff}}^2  \equiv \sigma _0^2  + \sigma _{{\rm{cw}}}^2
= \sigma _0^2  + \frac{{k_B T}}{2\pi \gamma} \ln \left(
\frac{q_{\max} }{q_{\min}}\right),
\label{For-sigma}
\end{equation}
where $\sigma_{\rm{eff}}$ depends on $q_z$ and $\Delta\beta$ from
the relation $q_{\min}=q_z\Delta\beta/2$.

It has been suggested that the intrinsic roughness $\sigma_0$ scales with molecular size, and $q_{\max}$ in Eq.\ (\ref{For-sigma}) has been usually estimated from the molecular size $R$ such that $q_{\max} = \pi/R$\cite{Braslau1988}.  However, studies of pure water and pure ethanol\cite{Schwartz1990,Sanyal1991} found $\sigma_0=0$, implying a perfectly uniform electron-density with no atomic or molecular discreteness of the liquid. It should be noted that in the studies mentioned above the contributions of $\sigma_0$  and $q_{\max}$ to the surface roughness according to Eq.\ (\ref{For-sigma}) could not be decoupled\cite{comment1}.  Subsequently, in a study of long chain C20 and C36 alkanes, by varying the liquid temperature, it was found that $\sigma_0 = 1.1$ {\AA}, significantly smaller than molecular size\cite{Ocko1994}.

In the present study we employ synchrotron X-ray
reflectivity studies to determine the relevant parameters of capillary wave theory, i.e., $\sigma_0$, $q_{\min}$, and $q_{\max}$, by systematically varying the surface tension of
simple alcohols (methanol, ethanol and propanol) and their
mixtures with water. The alcohol molecules are very close in size
to a water molecule (in particular methanol) thus minimizing the
presumed differences due to molecular size through
$\sigma_0$ or $q_{\max}$, yet allowing the continuous variation of
surface tension over a wide range (22 - 73 mN/m at room temperature) by changing the mixture concentration \cite{Sung2005}.
\section{Experimental Details}
The alcohols (methanol 99.9\%; ethanol 99.5\%; and 1-propanol
99.5\%), purchased from Fisher Chemicals, were used without
further purification. Ultrapure water (Millipore, Milli-Q, and
NANOpure, Barnstead; resistivity, 18.1 M$\Omega$cm) was used to
make the mixtures, without any buffer to adjust the pH (pH $\sim
6.5$). The surface tension of all solutions was measured at
21$^o$C using a DuNuoy Tensiometer. X-ray
reflectivity measurements were conducted on the Ames Laboratory
Liquid Surface Diffractometer, in Sector 6 of the Advanced
Photon Source (APS) at Argonne National Laboratory \cite{Vaknin2003,Bu2006}.  The highly monochromatic beam (16.2 keV was used for all liquids and mixtures and 8 keV for water/propanol mixtures) selected by a downstream Si double crystal
monochromator, is deflected onto the liquid surface to
a desired angle of incidence ($\alpha$) with respect to the
liquid surface by a second monochromator located on the
diffractometer. The trough containing the liquid samples ($\approx$10 cm
in diameter and $\sim 200 \mu$m deep) is enclosed
inside a temperature controlled and gas-tight aluminum canister.
A thin liquid film approximately $200 - 300$ $\mu$m deep, an active vibration isolation unit (JAS Mod-2), and a three-second waiting-time after a movement of any component of the diffractometer before counting photons are used to obviate the effect of undesirable mechanical agitations of the liquid.  Before the start of each measurement the shape (peak-height and width) of the totally reflected beam below the critical angle is confirmed to be practically the same as that of the direct beam\cite{Vaknin2003}.  The volume enclosing the trough is constantly purged with helium
bubbled through the corresponding water/alcohol mixture.

\section{Results and Discussion}
As shown in Eq.\ (\ref{For-sigma}), the effective surface roughness depends on the
instrumental resolution via $q_{\min}$.  To validate the resolution of our experimental set-up, we
conducted X-ray reflectivity (XR) from pure methanol under two
detector slit conditions, as shown in  Figure\ \ref{Ref0}(a). The
0.5 and 1.5 mm detector slit widths at a distance 756 mm from the
sample-center yield $\Delta\beta \approx 0.00066$ and 0.0020
radians, respectively.  The non-linear least-square fit to the data
using Eq.\ (\ref{For-RRF}) (solid lines) with the logarithmic
dependence on $q_z$  yields $\sigma_{\rm{eff}} =
4.9^{+0.1}_{-0.2}$ (for 0.5-mm slit) and $4.55^{+0.13}_{-0.14}$
{\AA} (for 1.5-mm slit) at $q_z=0.3$ {\AA}$^{-1}$. Detailed
analysis shows that $\sigma_0$ can take values from
$\sigma_0=0$ to $\sigma_0 = \sigma_{\rm{eff}}$ by allowing
$q_{\max}$ to vary without any constraints\cite{comment1}. In
addition, as pointed out in Ref.\ \cite{Pershan2000} the
logarithmic dependence of $\sigma_{\rm{eff}}$ on $q_z$ is so weak
that using a constant $\sigma^{\prime}_{\rm{eff}}$ yields similar
results as for $q_z$-dependent $\sigma_{\rm{eff}}$ (at a midpoint in the measured $q_z$-range; $q_z\approx 0.3$ {\AA}$^{-1}$).  Indeed,  this simpler approach gives within uncertainties the same result ($\sigma^{\prime}_{eff} = 4.90\pm0.13$ and $4.59\pm0.13$ {\AA} (for the 0.5 and 1.5 mm slits, respectively) as the one that assumes $q_z-$dependent $\sigma_{\rm{eff}}$.

To examine further resolution effects and the contribution of
capillary-waves, we measured X-ray diffuse scattering (XDS) for
water and methanol under the same slit conditions used to obtain
XR data (detector slit at 1.5 mm).  In the
past, the XDS from water and other liquids
\cite{Schwartz1990,Sanyal1991,Pershan2000,Shpyrko2004,Fukuto2006}
has been thoroughly investigated confirming capillary wave
predictions.  However, similar to specular reflectivity data, only $\sigma_{eff}$ can be determined whereas the values of $\sigma_0$ and $q_{\max}$ can be decoupled only if the temperature and/or surface tension are varied.  To a good approximation our resolution function is
rectangular (incident-beam-slit $\approx$0.08 mm and
beam-divergence $\sim 10^{-5}$ rad.).  Following the detailed
procedure in Refs.\ \cite{Fukuto2006,Sinha1988}, and by
defining $I(q_y,q_z) =
I_m(q_y,q_z)\sin\alpha/|T(\alpha)|^2|T(\beta)|^2 $ where the $I_m$
is measured intensity and $T(\alpha), T(\beta)$ are the
transmission functions, we find the normalized XDS is
 \cite{BuTobepublished}
\begin{equation}
\frac{I(q_{y},q_{z})}{I(0,q_{z})} = Cg(\alpha,\beta)+
(1-Cg_0)\hspace{0.2cm}\times \hspace{2.3cm} \label{For-XDS}
\end{equation}
\begin{equation}
 \frac{(\Delta q_{y}-2q_{y})\left|\Delta q_{y}-2q_{y}\right|^{\eta-1}+(\Delta q_{y}+2q_{y})\left|\Delta q_{y}+2q_{y}\right|^{\eta-1}}{2(q_{z}\Delta\beta/2)^{\eta}} \nonumber,
\end{equation}
where, $\Delta q_{y}=(q_{z}^{2}-2k_{0}q_{y})/(2q_{z}\Delta\beta)$,
$g(\alpha,\beta)=D(\alpha)D(\beta)/(D(\alpha)+D(\beta))$ ($D$ is
the x-ray penetration depth), and $g_0=g(\alpha,\alpha)$ accounts
for scattering from the bulk structure factor, and $C$ is a
constant determined experimentally at a small azimuthal angle
away from the scattering plane for specular reflectivity
condition \cite{Fukuto2006,BuTobepublished}. Figure\
\ref{Ref0}(b) shows XDS measurements for water and methanol at
fixed $q_z$ (i.e., varying $q_y = k_0(\cos\beta-\cos\alpha)$ by
changing $\alpha$ and $\beta$ such that
$\sin(\alpha)+\sin(\beta)=q_z/2k_0$ is constant) normalized to the measured reflectivity at $q_y=0$.  By normalizing the XDS data, the two coupled parameters, $\sigma_0$ and $q_{\max}$, are eliminated \cite{comment1}. The calculated intensity to the normalized data (solid lines) using Eq.\ (\ref{For-XDS}) with no adjustable parameters confirms the XDS are accounted for by capillary waves and also confirms the instrumental $q_{\min}$ value.

\begin{figure}[th1]
\includegraphics[width=3.0 in, angle=0]{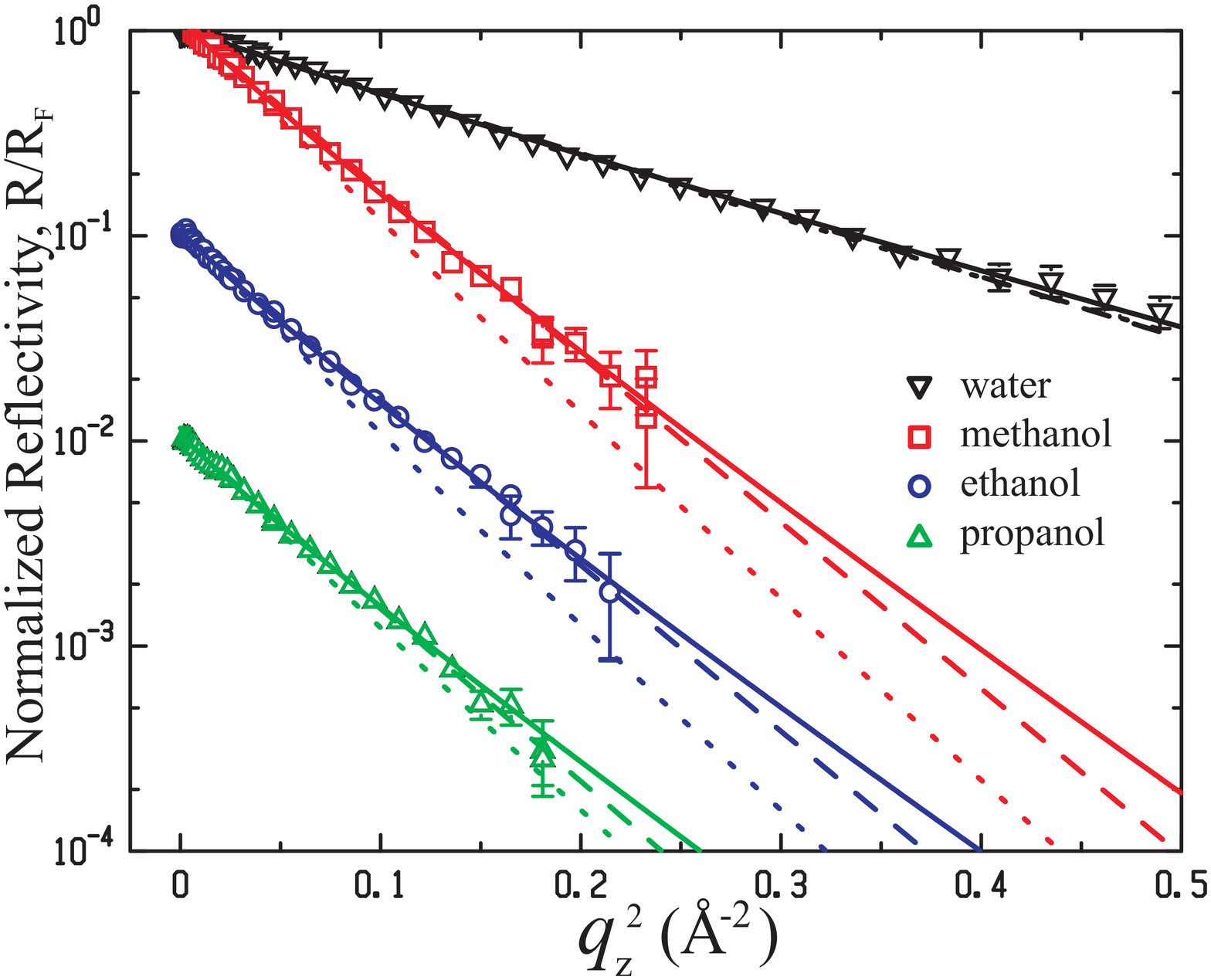}
\caption{\label{Ref1} (color online) Normalized XR of the pure liquids {\it
versus} $q_z^2$. The dotted lines are the best fits using Eq.\
(\ref{For-RRF}) with $q_{\min}=q_z\Delta\beta/2$ and a fixed
$q_{\max} = \pi/R$ varying only $\sigma_0$, and the solid lines
are obtained by varying $q_{\max}$. The dashed lines are the best
fits assuming a single parameter $q_z$-independent
$\sigma^\prime_{\rm{eff}}$.}
\end{figure}

Figure\ \ref{Ref1} shows the normalized reflectivities of the
pure liquids as indicated. Whereas the reflectivity of
water is significantly different, the reflectivities of all three
alcohols are hardly distinguishable.  Since surface tension of the
alcohols are very close in value, $\sigma_{\rm{cw}}$ is
practically the same for all three, indicating the intrinsic roughness ($\sigma_0$) is not dominated by molecular size. Three different procedures for analyzing the data in fig.\
\ref{Ref1} were examined. In the first (dotted lines), we assume that the only
free parameter in Eq.\ (\ref{For-RRF}) is $\sigma_0$, and fixed
$q_{\max} =\pi/R$ where $R$ is the average radius of each molecule; 1.93, 2.52, 2.85, and 3.1
{\AA} for water, methanol, ethanol, and propanol, respectively.
All the fits (dotted lines) yielded $\sigma_0 \approx 0$, and
except for water the fit is very poor. By relaxing the constraint
on $q_{{\max}}$ we get a better fit to the data (solid lines)
with a finite $\sigma_0$. However, as implied above
\cite{comment1}, the two parameters $\sigma_0$ and $q_{\rm{max}}$
are strongly coupled, and either the temperature\cite{Ocko1994} or the surface
tension of the system have to be changed to obtain $\sigma_0$ and $q_{\max}$.
By making an assumption (based on
the observation and discussion above) that all liquids have
within uncertainty similar $\sigma_0$ and $q_{\max}$, and by
fitting all data sets simultaneously, we obtain $\sigma_0=1.4 \pm
0.4$ {\AA} and $q_{\max}=0.152^{+0.1}_{-0.06}$ {\AA}$^{-1}$.
Clearly the optimal $q_{\max}$ value from
the fitting is significantly smaller than the value determined by
molecule-size $R$ ($q_{\max} \approx$ 1.63 and 1.01
{\AA}$^{-1}$ for water and propanol, for example). This suggests capillary wave theory breaks down at
an average length scale $l_r \approx \pi/q_{\max} =
21^{+13.2}_{-8.2}$ {\AA}.
A third way to fit the data is to simply assume
$q_z$-independent $\sigma^{\prime}_{\rm{eff}}$ as those shown by
dashed lines in fig.\ \ref{Ref1} demonstrating the
$q_z$-dependence of $\sigma_{\rm{eff}}$ is insignificant
\cite{Pershan2000}.

\begin{figure}[thl]\includegraphics[width=3.0 in, angle=0]{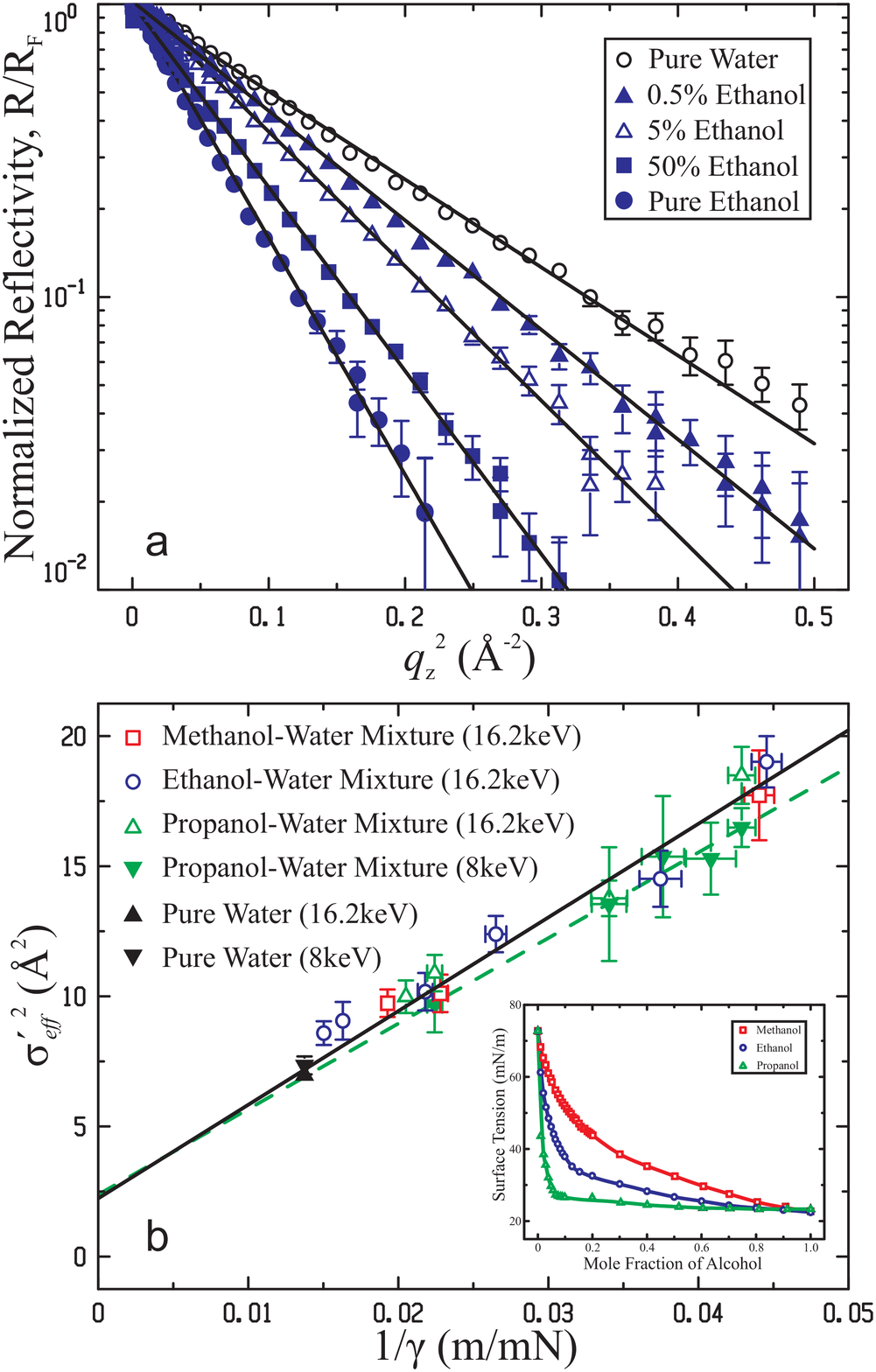}
\caption{\label{sigma_V_st} (color online) (a) Normalized X-ray reflectivity of
the ethanol-water mixtures versus $q_z^2$.  The solid lines are
linear fits to the (logarithm of) data yielding
$\sigma^{\prime}_{\rm{eff}}$. (b) Compilation of $\sigma_{\rm{eff}}^{\prime 2}$
values obtained from the analysis of the reflectivities versus
the inverse of the surface tension. The solid line (16.2 keV) and dashed line (8 keV) are the best
linear fits to the data. The inset in (b) shows the surface tension of
water-alcohol mixtures with varying bulk concentrations.}
\end{figure}

Figure\ \ref{sigma_V_st} (a) shows the normalized reflectivities
from the water/ethanol mixtures. Reflectivities from the
mixtures of methanol and propanol and their mixtures (not shown) are similar. The
solid lines are the best fits to the data using Eq.\
(\ref{For-RRF}) assuming a single parameter $\sigma^{\prime}_{\rm{eff}}$
($q_z$-independent). For all liquids, we carefully examined the region of the reflectivity near the critical angle,  and found excellent agreement with the average electron density of each mixture. We considered the atomic/molecular structure factor and found it has negligible effect on the extracted parameters. Also, we could not find any evidence of surface layering in the mixtures\cite{Sung2005}.

Figure\ \ref{sigma_V_st}(b) shows a compilation of all measured
effective surface roughness values ($\sigma_{\rm{eff}}^{\prime 2}$)
{\it versus} the inverse of surface tension for each mixture, as indicated. The
surface tension values of the mixtures are shown in the inset
\cite{Sung2005}. To a good approximation, all the measured
roughness values fall on a linear curve (solid line) that within
experimental error confirms Eq.\ (\ref{For-sigma}). This behavior
strongly suggests the profile of the pure and liquid mixtures of
small molecules is {\it predominantly} determined by surface
tension (at a given temperature).  From fig.\ \ref{sigma_V_st}(b) we
obtain values for $\sigma_0$ for the slope $\Delta = k_BT\ln(q_{\max}/q_{\min})/2\pi$ as listed in table 1.
The values of $\sigma_0^2$, for both 8 and 16.2 keV, are consistent within uncertainties.  The larger difference, between the values of $\Delta$, is mainly due to different detector slit configurations used at the two X-ray energies, yielding two different values for $q_{\min}$.
In the following, we make the assumption that $\sigma_0$ and $q_{max}$ are, within uncertainties, the same for all liquids and mixtures.  Although they should differ for the various liquids, the variation is most likely smaller than the experimental uncertainty.  This is rationalized based on the results of the pure alcohols above that yield within error the same $\sigma_0$ and $q_{max}$, although the three alcohols differ significantly in molecular size.
\begin{table}[!h]
\caption{\label{tab1}}
\begin{tabular}{ccccc}
\hline\hline
&&&&\tabularnewline
XR Energy&
$\sigma_{0}^{2}$ &
$\Delta$ &
$q_{\max}$ &
$l_{r}$\tabularnewline
(keV) &
({\AA}$^2$) &
(x10$^{-23}$ J)&
({\AA}$^{-1})$&
\tabularnewline
&&&&\tabularnewline
\hline
&&&&\tabularnewline
16.2 &
$2.2_{-0.2}^{+0.3}$&
$360_{-34}^{+42}$&
$0.08_{-0.03}^{+0.07}$&
$40_{-20}^{+28}$\tabularnewline

&&&&\tabularnewline
8 &
$2.4_{-0.3}^{+0.3}$&
$329_{-30}^{+35}$&
$0.07_{-0.03}^{+0.05}$&
$48_{-20}^{+29}$\tabularnewline
\hline\hline
\end{tabular}
\end{table}

The value of the intrinsic roughness $\sigma_0 = 1.5 \pm 0.2$
{\AA} is very close to that of bond lengths in our systems (e.g.,
C-C and C-O with bond lengths 1.54 and 1.43 {\AA},
respectively).  The theory for X-ray reflectivity above assumes
the electron density is a continuum, but physically, the electrons
are concentrated around discrete nuclei thereby giving rise to
intrinsic roughness on the scale of atomic separations, hypothetically, even at
zero temperature where all thermal motions are frozen. This is
in agreement with the study of liquid alkanes, where it was found
that $\sigma_0 = 1.1$ {\AA} and  correctly associated with
interatomic C-C bond length\cite{Ocko1994}. Although in the present study of simple liquids, we find $\sigma_0$ is on the order of average interatomic distances, we do not rule out that in more complex liquids, the intrinsic roughness may depend on molecular or aggregate size.

From the slope of the curve $\Delta$ in fig.\ \ref{sigma_V_st}(b)
we can estimate the value of $q_{\max}$ assuming $q_{\min} =
\langle q_z\rangle\Delta\beta/2$ where $\langle q_z \rangle =
0.3$ {\AA}$^{-1}$, at about the midpoint of the range over which
the reflectivities were measured, as listed in Table\ \ref{tab1} (the slopes of the curves at 8 and 16.2 keV are slightly different due to different slit configuration).  The corresponding $l_r$ lengths
are in agreement with the values obtained from the analysis of the pure liquids above.
Our view is that the value of $l_r$ reflects bulk correlation length, a length scale below which capillary wave theory breaks down. Although thermal excitations may exist for $q \sim \pi/R$ and even at larger values, due to inter-molecular vibrational states, these are of a different nature than those of capillary-waves.  Due bulk short-range-order, these may be more like optical phonons in solids, where it is common that their average amplitude (Debye-Waller factor) is only a few percents of interatomic distances thus negligibly contributing compared to capillary-waves. This is in agreement with the current view of capillary-wave theories and simulations that assume fluctuations of the interface on length scales larger than the bulk correlation length \cite{Mecke1999}.  The influence of the bulk properties on the surface behavior has been discussed theoretically and experimentally in the context of power spectra fluctuations  of a liquid surface\cite{Bouchiat1971}.

\section{Summary}
In the present synchrotron X-ray study, we systematically
determined the profiles of the liquid/vapor interfaces of water
and its mixtures with methanol, ethanol and propanol at a fixed
temperature, with the premise that while surface tension changes
by mixing water with alcohols, the molecular size at the
interface is as close as possible to that of the water
molecule. We emphasize that the values we report for $\sigma_0$ and $q_{max}$, although expected to vary among the different liquids, are approximate, and should be considered as a lower and an upper limits, respectively.  We find that the intrinsic roughness for these simple
liquids reflects inter-atomic distances setting a low limit to
$\sigma_0$.  We also find that the upper wave-vector cutoff for
capillary waves (although with inherently large uncertainty) is appreciably smaller than that expected by assuming it is dominated by molecular diameter.  This implies the break-down of capillary wave theory at a few
molecular diameters due to {\it rigidity} of the surface membrane over short length scales on the order of bulk correlation lengths.

\acknowledgments We thank D. S. Robinson for technical support at the 6-ID
beamline and A. Travesset, B. M. Ocko, and M. Fukuto for helpful discussions.
The MUCAT sector at the APS is supported by the U.S.
DOE, Basic Energy Sciences, Office of Science, through Ames
Laboratory under contract under Contract No. DE-AC02-07CH11358. Use of the Advanced
Photon Source is supported by the U.S. DOE, Basic Energy
Sciences, Office of Science, under Contract No. W-31-109-Eng-38.
D. K. acknowledges support from the Quantum Photonic Science
Research Center (SRC) at Hanyang University.



\begin{references}
\bibitem{comment1} Equation\ (\ref{For-sigma}) can be written as follows $\sigma_{\rm{eff}}^2 =\left[\sigma_0^2+\frac{{k_BT}}{2\pi \gamma}
 \ln \left(q_{\max}\right)\right]-\frac{{k_B T}}{2\pi \gamma} \ln \left({q_z\Delta\beta/2 }\right)$
 showing the contributions of $\sigma_0$ and $q_{\max}$ cannot be decoupled by varying slits size, but only by varying temperature and/or surface tension.
\end{references}
\end{document}